\documentstyle[psbox]{WORKSHOP}
\begin{document}

\title{
Suzaku Observation of HESS~J1507$-$622 \\ 
}

\author{
H. Matsumoto$^1$, Y. Terada$^2$, A. Bamba$^3$,\\ 
M. Sakai$^4$, O. Tibolla$^5$ and S. Kaufmann$^6$
\\[12pt]  
%
$^1$  Kobayashi-Maskawa Institute (KMI), Nagoya University, Furo-cho, Chikusa-ku, Nagoya, Aichi, Japan 464-8602\\
$^2$  Graduate School of Science and Engineering, Saitama University, Japan \\
$^3$  Department of Physics and Mathematics, Aoyama Gakuin University, Japan \\
$^4$ Japan Aerospace Exploration Agency, Japan \\
$^5$ Universit\"at W\"urzburg, Germany \\
$^6$ Landessternwarte, Universit\"at Heidelberg, Germany \\
%
{\it E-mail(HM): matumoto@u.phys.nagoya-u.ac.jp} 
}

\abst{
HESS~J1507$-$622 is one of the bright unidentified TeV
objects. HESS~J1507$-$622 is unique, since the location of
the object is off the Galactic disk.  We observed the
HESS~J1507$-$622 region with the Suzaku XIS, and found no
obvious counterpart although there is no severe interstellar
extinction. However, there are two interesting X-ray
objects; SRC1 is a bright extended source, and SRC2 is a
faint diffuse object. If either of them is a counterpart,
the flux ratio between TeV and X-ray is large, and
HESS~J1507$-$622 is a real dark particle accelerator.
}

\kword{acceleration of particles -- X-rays: individual (HESS J1507$-$622)}

\maketitle
\thispagestyle{empty}

\section{Introduction}

HESS~J1507$-$622 (hereafter, HESSJ1507) is one of the
bright ($\sim$8\% of the Crab) unidentified TeV objects
(H.E.S.S.~Collaboration et al. 2011).  HESSJ1507 is unique
since it lies $\sim$3.5~deg from the Galactic plane, while
most of the unidentified objects are located within
$\pm$1~deg from the Galactic plane.  The TeV spectrum can be
described by a power-law model with $\Gamma = 2.24 \pm 0.16
(\mbox{stat}) \pm 0.20 (\mbox{sys})$, and a flux of
$F(1-10~\mbox{TeV}) = 5.1 \times
10^{-12}$~erg~cm$^{-2}$~s$^{-1}$ (H.E.S.S.~Collaboration
et al. 2011). 

HESSJ1507 is also detected by Fermi-LAT (Domainko \& Ohm
2012).  The GeV spectrum can be described by a power-law
model with $\Gamma = 1.7 \pm 0.1 (\mbox{stat}) \pm 0.2
(\mbox{sys})$.  The flux is $F(0.3-300~{\mbox{GeV}}) = 3.5
  \times 10^{-11}$~erg~cm$^{-2}$~s$^{-1}$ (Domainko \& Ohm
  2012).

%
%
\begin{figure*}
\centering
\psbox[xsize=0.65\textwidth]{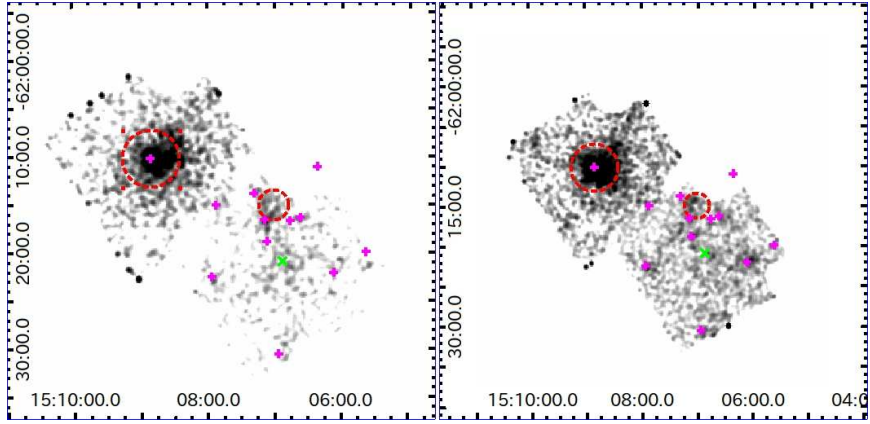}
\caption{Suzaku XIS3 images of HESSJ1507: (left) 0.4--2 keV,
  (right) 2--10keV.  The position of the TeV gamma-ray
  emission of HESSJ1507 is shown as a cross mark. X-ray
  sources detected by Chandra are shown as plus
  marks. Spectral regions of SRC1 and SRC2 are shown as
  broken circles.}
\end{figure*}

\section{Suzaku Observation}

Suzaku observed two regions around HESSJ1507; the center
region and the off-center region $\sim$ 15~arcmin northeast
from the center. The exposure time is 79.9~ks for the center
region and 40.9~ks for the off-center region.

\section{Results}

\section{X-ray image}

The XIS3 images are shown in Fig.~1. The position of
HESSJ1507 is shown as a cross mark. There is no X-ray source
which coincides with the peak of the TeV emission.  The same
regions were observed by Chandra, and many objects were
found (H.E.S.S.~Collaboration et al. 2011); most of them
were identified as stars. These Chandra sources are shown as
plus marks. Many of them are also detected by Suzaku.

There is a conspicuous source at $(\alpha,
\delta)_{\mbox{J2000}} = (\mbox{15:08:51},~\mbox{−62:10:18})$. We call
this source SRC1.  SRC1 was also detected with Chandra and
ROSAT (H.E.S.S.~Collaboration et al. 2011).  Though SRC1 is
outside of the 3σ significance contour of the TeV emission
(H.E.S.S.~Collaboration et al. 2011), we will study the
X-ray characteristics of this object in the following.

Chandra discovered a faint diffuse object, which is also
detected by Suzaku; this source is shown by the small broken
circle in Fig.~1, and we call this source SRC2.

\subsection{SRC1}

Fig.~2 shows a radial profile of SRC1 in the 3--10 keV band
obtained from the XIS FI data (XIS0+3). The profile cannot
be described by a point spread function, and SRC1 is a
spatially extended object.

Time variability of SRC1 was examined. The light curve of
SRC1 in the 2--10 keV band extracted from XIS3 is shown in
Fig.~3. There is no significant time variability. No
variability can be seen also in the 0.4--10~keV and
0.4--2~keV bands.

\begin{figure}[h]
\begin{minipage}[t]{0.24\textwidth}
\centering
\psbox[xsize=\textwidth]{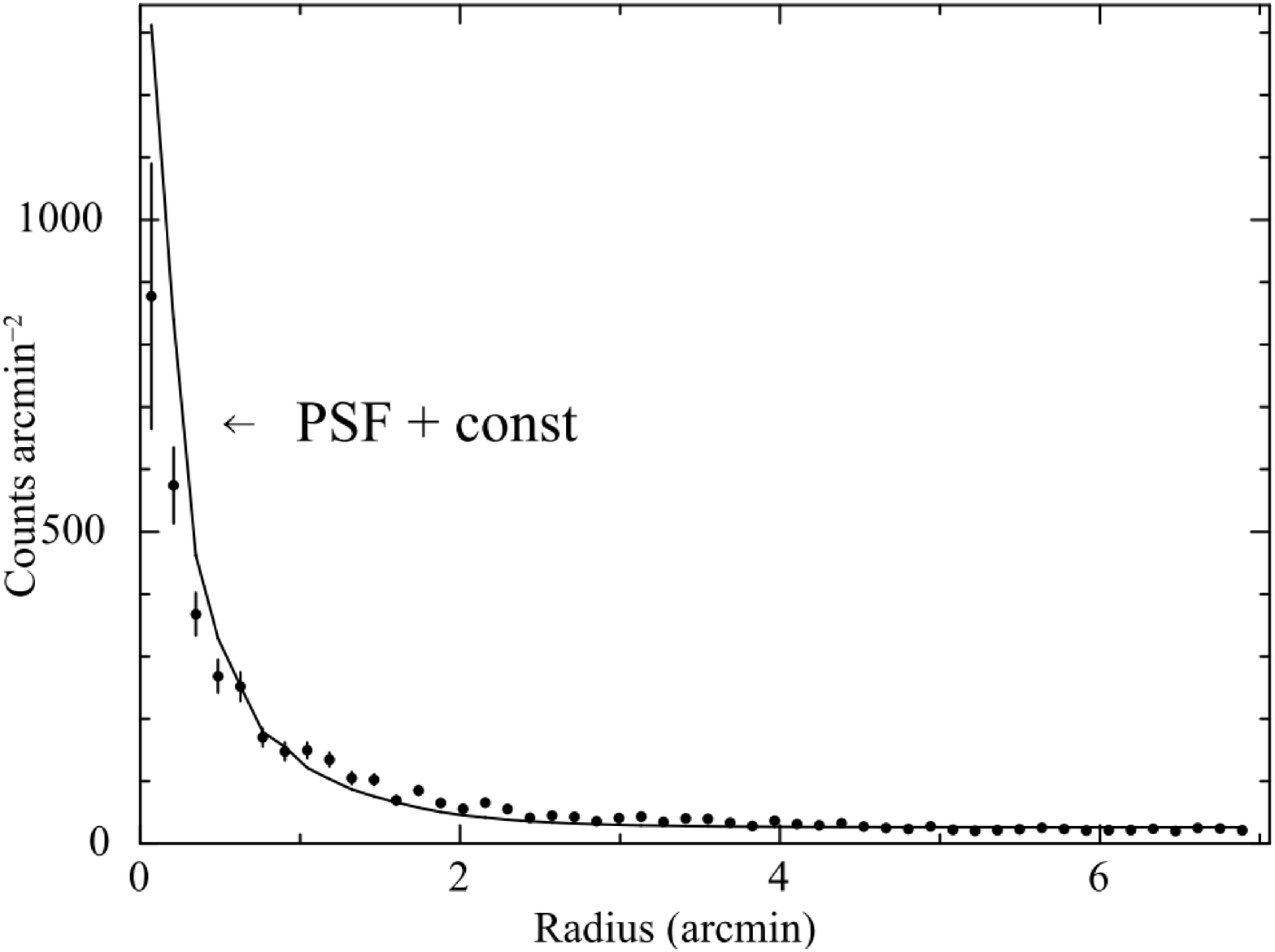}
\caption{Radial profile of SRC1 in the 3--10 keV band from
  the XIS FI sensor (XIS0+3).}
\end{minipage}
\begin{minipage}[t]{0.24\textwidth}
\centering
\psbox[xsize=\textwidth]{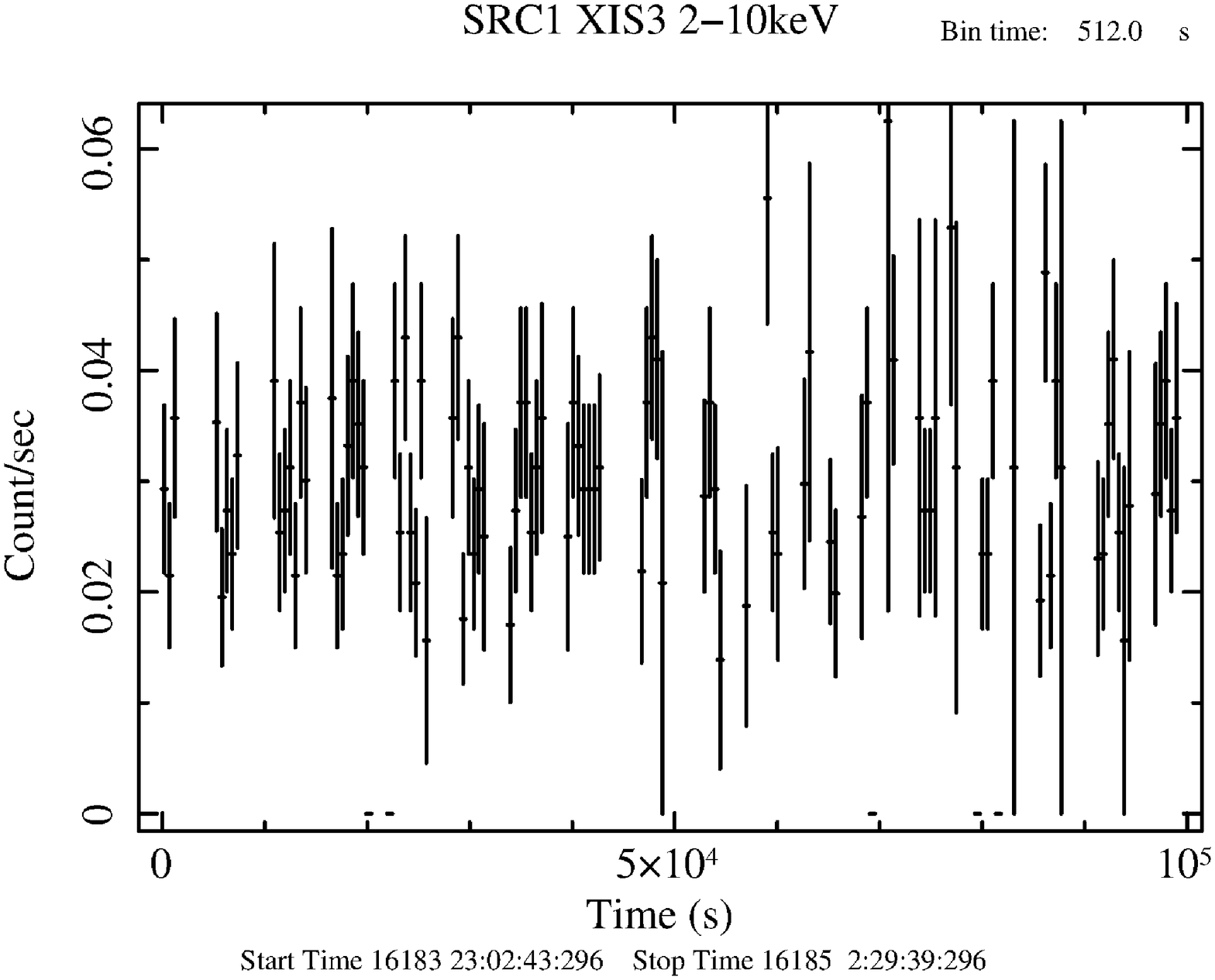}
\caption{Radial profile of SRC1 in the 3--10 keV band from
  the XIS FI sensor (XIS0+3).}
\end{minipage}
\end{figure}

The X-ray spectrum is shown in the left panel of Fig.~4. The
spectrum can be described by an absorbed power-law
model. The best-fit parameters are $N_{\rm H} = 0.5 \pm 0.1
\times 10^{22}$~ cm$^{-2}$, $\Gamma = 1.8 \pm 0.1$. The
flux is $F(2-10~\mbox{keV}) =9.7 \times
10^{-13}$~erg~s$^{-1}$~cm$^{-2}$ . These parameters are
compatible with those determined by Chandra
(H.E.S.S.~Collaboration et al. 2011).

\begin{figure}[h]
\centering
\psbox[xsize=0.3\textwidth]{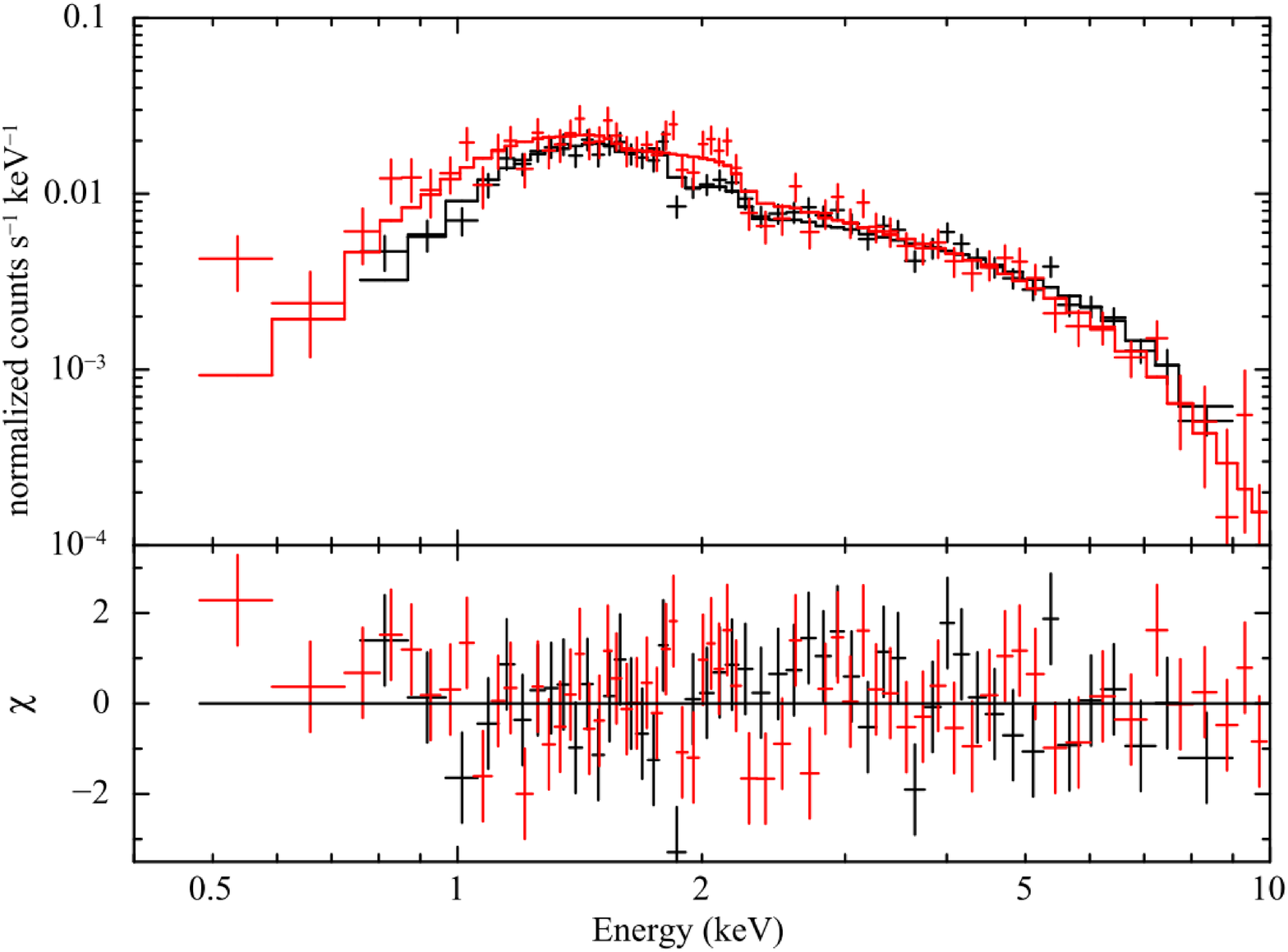}
\caption{X-ray spectrum of SRC1. Black and red
data points represent those of the FI (XIS0+3) and BI (XIS1) sensors,
respectively. The best-fit power low model is shown as solid lines.
}
\end{figure}

\vspace{-.5em}

\subsection{SRC2}

The X-ray spectrum of SRC2 is shown in Fig.~5. The spectrum
can be described by a power-law model. The best-fit
parameters are $N_{\rm H} = (0.8 \pm 0.5) \times
10^{22}$~cm$^{-2}$, $\Gamma = 2.4 \pm 0.5$. 
The flux in the 2--10~keV band is $F(2-10~\mbox{keV})
=0.8 \times 10^{-13}$~erg~s~$^{-1}$~cm$^{-2}$.

\begin{figure}[h]
\centering
\psbox[xsize=0.3\textwidth]{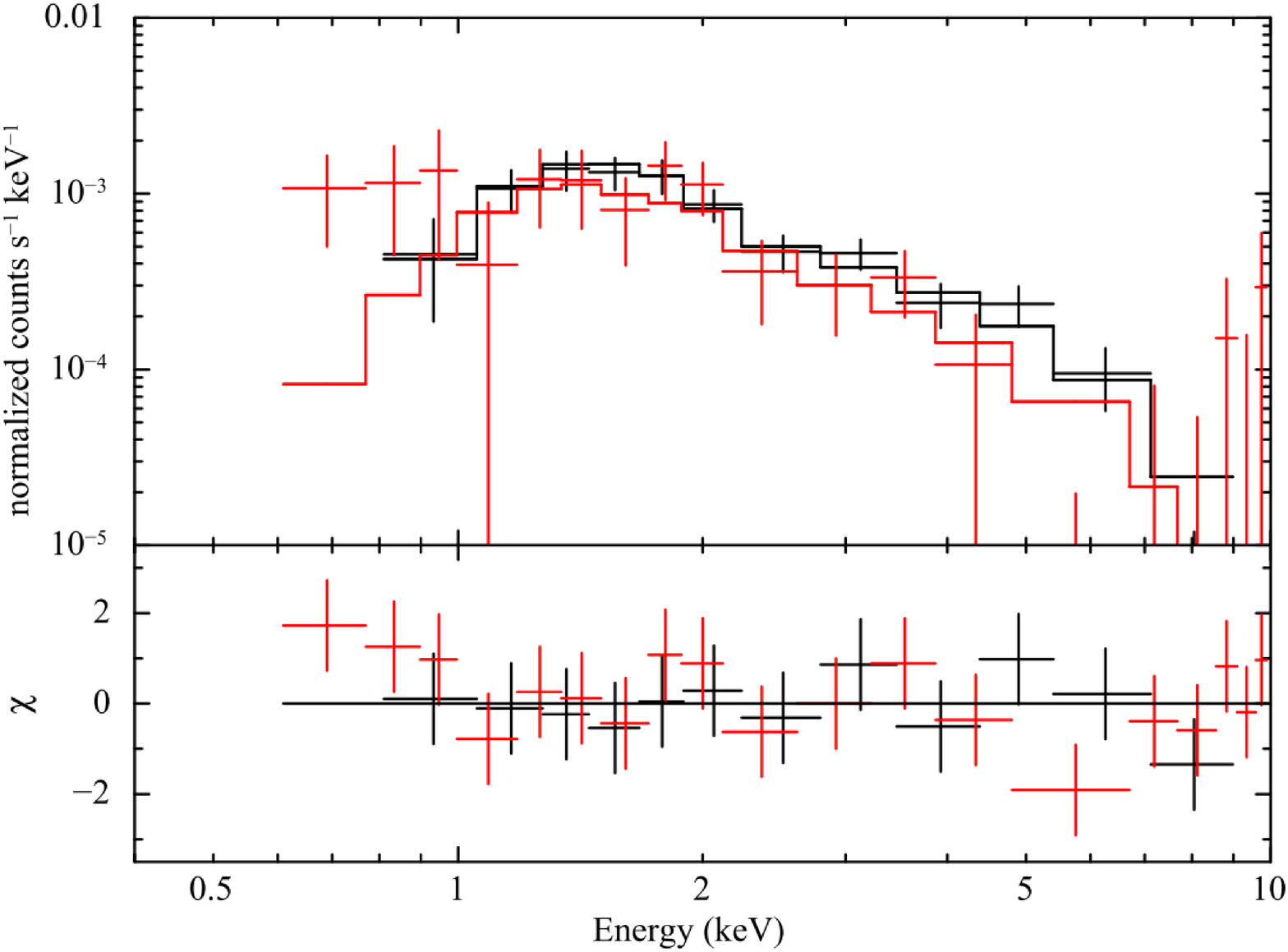}
\caption{X-ray spectrum of SRC2. Black and red
data points represent those of the FI (XIS0+3) and BI (XIS1) sensors,
respectively. The best-fit power low model is shown as solid lines.
}
\end{figure}

\vspace{-1em}

\section{Discussion}

Although the location of HESSJ1507 is away from the Galactic
disk and the X-ray observation does not suffer from a heavy
interstellar absorption, there is no obvious X-ray
counterpart to HESSJ1507. If SRC1 is an X-ray counterpart to
HESSJ1507, the flux ratio
$F(1-10~\mbox{TeV})/F(2-10~\mbox{keV})$ would be $\sim$5.3.
If SRC2 is a counterpart, the ratio would be $\sim$64.
H.E.S.S.~Collaboration et al. (2011) suggest that SRC2 could
be a bright part of a larger and fainter source. We can see
no hint in the Suzaku data suggesting that possibility.
Since the total Galactic HI column density is $N_{\rm H} =
5.1 \times 10^{21}$~cm$^{-2}$, both SRC1 and SRC2 can be
thought as a Galactic object. If the density of interstellar
matter is assumed to be 1~cm$^{-3}$, the distance to SRC1
is $\sim (1.7 \pm 0.2)$~kpc, and that to SRC2 is $\sim (2.6
\pm 1.7)$~kpc.

\vspace{-.5em}

\section*{References}

\re
Domainko, W. and Ohm, S. 2012, A\&A, 545, A94

\re
H.E.S.S.~Collaboration, Acero, F., Aharonian, F., et al.\ 2011, A\&A, 525, A45

\label{last}

\end{document}